\documentclass[11pt]{article}%
\usepackage{amsmath}
\usepackage{graphicx}
\usepackage{amsfonts}
\usepackage{amssymb}
\usepackage{fullpage}%
\newtheorem{theorem}{Theorem}

\newtheorem{lemma}[theorem]{Lemma}

\newtheorem{proposition}[theorem]{Proposition}

\newenvironment{proof}[1][Proof]{\textbf{#1.} }{\ \rule{0.5em}{0.5em}}
\begin{document}

\title{Quantum Computing and Hidden Variables II: The Complexity of Sampling Histories}
\author{Scott Aaronson\thanks{University of California, Berkeley. \ Email:
aaronson@cs.berkeley.edu.}}
\date{}
\maketitle

\begin{abstract}
This paper shows that, if we could examine the entire history of a hidden
variable, then we could efficiently solve problems that are believed to be
intractable even for quantum computers. \ In particular, under any
hidden-variable theory satisfying a reasonable axiom called \textquotedblleft
indifference to the identity,\textquotedblright\ we could solve the Graph
Isomorphism and Approximate Shortest Vector problems in polynomial time, as
well as an oracle problem that is known to require quantum exponential time.
\ We could also search an $N$-item database using $O\left(  N^{1/3}\right)  $
queries, as opposed to $O\left(  N^{1/2}\right)  $ queries with Grover's
search algorithm. \ On the other hand, the $N^{1/3}$ bound is optimal, meaning
that we could probably \textit{not} solve $\mathsf{NP}$-complete problems in
polynomial time. \ We thus obtain the first good example of a model of
computation that appears \textit{slightly} more powerful than the quantum
computing model.

\end{abstract}

\section{Introduction\label{INTRO}}

It is often stressed that hidden-variable theories, such as Bohmian mechanics,
yield exactly the same predictions as ordinary quantum mechanics. \ On the
other hand, these theories describe a different picture of physical reality,
with an additional layer of dynamics beyond that of a state vector evolving
unitarily. \ This paper addresses a question that, to our knowledge, had never
been raised before: \textit{what is the computational complexity of simulating
that additional dynamics?} \ In other words, if we could examine a hidden
variable's entire history, then could we solve problems in polynomial time
that are intractable even for quantum computers?

We present strong evidence that the answer is yes. \ The Graph Isomorphism
problem asks whether two graphs $G$ and $H$ are isomorphic; while given a
basis for a lattice $\mathcal{L}\in\mathbb{R}^{n}$, the Approximate Shortest
Vector problem asks for a nonzero vector in $\mathcal{L}$ within a $\sqrt{n}%
$\ factor of the shortest one. \ We show that both problems are efficiently
solvable by sampling a hidden variable's history, provided the hidden-variable
theory satisfies a reasonable axiom that we call \textquotedblleft
indifference to the identity operation.\textquotedblright\ \ By contrast,
despite a decade of effort, neither problem is known to lie in $\mathsf{BQP}$,
the class of problems solvable in quantum polynomial time with bounded error
probability.\footnote{See www.complexityzoo.com for more information about the
complexity classes mentioned in this paper.} \ Thus, if we let $\mathsf{DQP}%
$\ (Dynamical Quantum Polynomial-Time) be the class of problems solvable in
our new model, then this already provides circumstantial evidence that
$\mathsf{BQP}$\ is strictly contained in $\mathsf{DQP}$.

However, the evidence is stronger than this. \ For we actually show that
$\mathsf{DQP}$\ contains an entire \textit{class} of problems, of which Graph
Isomorphism\ and Approximate Shortest Vector\ are special cases.\ \ Computer
scientists know this class as \textit{Statistical Zero Knowledge}, or
$\mathsf{SZK}$. \ Furthermore, in previous work \cite{aarcol} we showed that
\textquotedblleft relative to an oracle,\textquotedblright\ $\mathsf{SZK}$\ is
not contained in $\mathsf{BQP}$. \ This is a technical concept implying that
any proof of$\ \mathsf{SZK}\subseteq\mathsf{BQP}$ would require techniques
unlike those that are currently known. \ Combining our result that
$\mathsf{SZK}\subseteq\mathsf{DQP}$ with the oracle separation of
\cite{aarcol}, we obtain that $\mathsf{BQP}\neq\mathsf{DQP}$\ relative to an
oracle as well. \ Given computer scientists' longstanding inability to
separate basic complexity classes, this\ is nearly the best evidence one could
hope for that sampling histories yields more power than standard quantum computation.

Besides solving $\mathsf{SZK}$\ problems, we also show that by sampling
histories, one could search an unordered database of\ $N$\ items for a single
\textquotedblleft marked item\textquotedblright\ using only $O\left(
N^{1/3}\right)  $\ database queries. \ By comparison, Grover's quantum search
algorithm \cite{grover}\ requires $\Theta\left(  N^{1/2}\right)  $ queries,
while classical algorithms require $\Theta\left(  N\right)  $
queries.\footnote{For readers unfamiliar with asymptotic notation: $O\left(
f\left(  N\right)  \right)  $\ means \textquotedblleft at most order $f\left(
N\right)  $,\textquotedblright\ $\Omega\left(  f\left(  N\right)  \right)
$\ means \textquotedblleft at least order $f\left(  N\right)  $%
,\textquotedblright\ and $\Theta\left(  f\left(  N\right)  \right)  $\ means
\textquotedblleft exactly order $f\left(  N\right)  $.\textquotedblright} \ On
the other hand, we also show that our $N^{1/3}$\ upper bound is the best
possible---so even in the histories model, one cannot search an $N$-item
database in $\left(  \log N\right)  ^{c}$\ steps for some fixed power $c$.
\ This implies that $\mathsf{NP}\not \subset \mathsf{DQP}$ relative to an
oracle, which in turn suggests that $\mathsf{DQP}$ is \textit{still} not
powerful enough to solve $\mathsf{NP}$-complete problems in polynomial time.
\ Note that while Graph Isomorphism\ and Approximate Shortest Vector\ are in
$\mathsf{NP}$, it is strongly believed that they are not $\mathsf{NP}$-complete.

At this point we should address a concern that many readers will have. \ Once
we extend quantum mechanics by positing the \textquotedblleft
unphysical\textquotedblright\ ability to sample histories, isn't it completely
unsurprising if we can then solve problems that were previously intractable?
\ We believe the answer is no, for three reasons.

First, almost every change that makes the quantum computing model more
powerful, seems to make it \textit{so much} more powerful that $\mathsf{NP}%
$-complete\ and even harder problems become solvable efficiently. \ To give
some examples, $\mathsf{NP}$-complete\ problems can be solved in polynomial
time using a nonlinear Schr\"{o}dinger equation, as shown by Abrams and Lloyd
\cite{al}; using closed timelike curves, as shown by Bacon \cite{bacon}; or
using a measurement rule of the form $\left\vert \psi\right\vert ^{p}$\ for
any $p\neq2$,\ as shown by us \cite{aarisl}. \ It is also easy to see that we
could solve $\mathsf{NP}$-complete\ problems\ if, given a quantum state
$\left\vert \psi\right\rangle $, we could request a classical description of
$\left\vert \psi\right\rangle $, such as a list of amplitudes or a preparation
procedure.\footnote{For as Abrams and Lloyd \cite{al}\ observed, we can so
arrange things that $\left\vert \psi\right\rangle =\left\vert 0\right\rangle
$\ if an $\mathsf{NP}$-complete instance of interest to us has no solution,
but\ $\left\vert \psi\right\rangle =\sqrt{1-\varepsilon}\left\vert
0\right\rangle +\sqrt{\varepsilon}\left\vert 1\right\rangle $\ for some tiny
$\varepsilon$\ if it has a solution.} \ By contrast, ours is the first
independently motivated model we know of that seems more powerful than quantum
computing, but only \textit{slightly} so.\footnote{One can define other, less
motivated, models with the same property by allowing \textquotedblleft
non-collapsing measurements\textquotedblright\ of quantum states,\ but these
models are very closely related to ours. \ Indeed, a key ingredient of our
results will be to show that certain kinds of non-collapsing measurements can
be \textit{simulated} using histories.} \ Moreover, the striking fact that
unordered search in our model takes about $N^{1/3}$\ steps, as compared to $N$
steps classically and $N^{1/2}$\ quantum-mechanically, suggests that
$\mathsf{DQP}$ somehow \textquotedblleft continues a
sequence\textquotedblright\ that begins with $\mathsf{P}$\ and $\mathsf{BQP}$.
\ It would be interesting to find a model in which search takes $N^{1/4}$\ or
$N^{1/5}$\ steps.

The second reason our results are surprising is that, given a hidden variable,
the distribution over its possible values at any \textit{single} time is
governed by standard quantum mechanics, and is therefore efficiently samplable
on a quantum computer. \ So if examining the variable's history confers any
extra computational power, then it can only be because of
\textit{correlations} between the variable's values at different times.

The third reason is our criterion for success. \ We are not saying merely that
one can solve Graph Isomorphism under \textit{some} hidden-variable theory; or
even that, under any theory satisfying the indifference axiom, there exists an
algorithm to solve it; but rather that there exists a \textit{single}
algorithm that solves Graph Isomorphism under any theory satisfying
indifference. \ Thus, we must consider even theories that are specifically
designed to thwart such an algorithm.

But what is the motivation for our results? \ The first motivation is that,
within the community of physicists who study hidden-variable theories such as
Bohmian mechanics, there is great interest in actually \textit{calculating}
the hidden-variable trajectories for specific physical systems \cite{pdh,gm}.
\ Our results show that, when many interacting particles are involved, this
task might be fundamentally intractable, even if a quantum computer is
available. \ The second motivation is that, in classical computer science,
studying \textquotedblleft unrealistic\textquotedblright\ models of
computation has often led to new insights into realistic ones;\ and likewise
we expect that the $\mathsf{DQP}$ model\ could lead to new results
about\ standard quantum computation. \ Indeed, in a sense this has already
happened. \ For our result that\ $\mathsf{SZK}\not \subset \mathsf{BQP}%
$\ relative to an oracle \cite{aarcol} grew out of work on the $\mathsf{BQP}$
versus $\mathsf{DQP}$ question. \ Yet the \textquotedblleft quantum lower
bound for the collision problem\textquotedblright\ underlying that result
provided the first evidence that cryptographic hash functions could be secure
against quantum attack, and ruled out a large class of possible quantum
algorithms for Graph Isomorphism, Approximate Shortest Vector, and related problems.

\subsection{Outline of Paper\label{OUTLINE}}

The precise definition of a hidden-variable theory\ that we use in this paper
was developed in a companion paper\ \cite{aaronson}. \ Familiarity with
\cite{aaronson}\ is helpful but not essential for understanding this paper.
\ In Section \ref{MODEL}, we review the relevant concepts from \cite{aaronson}%
, and then formally define $\mathsf{DQP}$\ as the class of problems solvable
by a classical polynomial-time algorithm with access to a \textquotedblleft
history oracle.\textquotedblright\ \ Given a sequence of quantum circuits as
input, this oracle returns a sample from a corresponding distribution over
histories of a hidden variable, according to some hidden-variable theory
$\mathcal{T}$. \ The oracle can choose $\mathcal{T}$\ \textquotedblleft
adversarially,\textquotedblright\ subject to two constraints: $\mathcal{T}%
$\ must be robust to small errors (since otherwise the definition of
$\mathsf{DQP}$\ could depend on the choice of gate set), and it must satisfy
the indifference axiom.

So what is the indifference axiom, then? \ Intuitively it says that, given a
bipartite state $\left\vert \psi\right\rangle \in\mathcal{H}_{A}%
\otimes\mathcal{H}_{B}$ (entangled or unentangled), if a unitary operation
acts only on the $\mathcal{H}_{A}$\ part of $\left\vert \psi\right\rangle
$\ (i.e. has the form $U\otimes I$), then the hidden-variable transitions can
also only involve the $\mathcal{H}_{A}$\ part. \ Note that this is quite
different from \textit{locality} in the sense of Bell's theorem: the
probability of transitioning between two basis states $\left\vert
x_{A}\right\rangle \otimes\left\vert x_{B}\right\rangle $\ and $\left\vert
y_{A}\right\rangle \otimes\left\vert x_{B}\right\rangle $\ can depend on the
complete state $\left\vert \psi\right\rangle $; all we require is that if
$x_{B}\neq y_{B}$, then the probability of transitioning between $\left\vert
x_{A}\right\rangle \otimes\left\vert x_{B}\right\rangle $\ and $\left\vert
y_{A}\right\rangle \otimes\left\vert y_{B}\right\rangle $ is zero.
\ Indifference is a substantive axiom, and is violated (for example) by
Bohmian mechanics. \ However, to us it simply expresses the idea that, if we
have a state such as $\left(  \left\vert a\right\rangle +\left\vert
b\right\rangle +\left\vert c\right\rangle +\left\vert d\right\rangle \right)
/2$, and a partial measurement yields a new state%
\[
\frac{\left\vert a\right\rangle +\left\vert b\right\rangle }{2}\left\vert
R_{ab}\right\rangle +\frac{\left\vert c\right\rangle +\left\vert
d\right\rangle }{2}\left\vert R_{cd}\right\rangle ,
\]
where $\left\vert R_{ab}\right\rangle $\ and $\left\vert R_{cd}\right\rangle
$\ denote two configurations of a recording apparatus, then so long as we
leave the recording apparatus alone, all further hidden-variable transitions
should be between $\left\vert a\right\rangle $\ and $\left\vert b\right\rangle
$\ or between $\left\vert c\right\rangle $\ and $\left\vert d\right\rangle $,
not between (say) $\left\vert a\right\rangle $ and $\left\vert c\right\rangle
$. \ If we abandoned this axiom, then we would need some other way to rule out
the degenerate hidden-variable theory, which takes the hidden-variable values
at different times to be completely independent of one another. \ Were this
\textquotedblleft product theory\textquotedblright\ allowed, we would have
$\mathsf{DQP}=\mathsf{B}\mathsf{QP}$\ for trivial reasons.

An earlier version of this paper required another axiom---\textit{symmetry}
under permutations of basis states---which seems much harder to justify than
indifference. \ However, we have since been able to eliminate the dependence
of our algorithms on the symmetry axiom.

Section \ref{RESULTS}\ establishes the most basic facts about $\mathsf{DQP}$:
for example, that $\mathsf{BQP}\subseteq\mathsf{DQP}$, and that $\mathsf{DQP}%
$\ is independent of the choice of gate set. \ Then Section \ref{JUGGLE}%
\ presents the \textquotedblleft juggle subroutine,\textquotedblright\ a
crucial ingredient in both main algorithms of the paper. \ Given a state of
the form $\left(  \left\vert a\right\rangle +\left\vert b\right\rangle
\right)  /\sqrt{2}$ or $\left(  \left\vert a\right\rangle -\left\vert
b\right\rangle \right)  /\sqrt{2}$, the goal of this subroutine is to
\textquotedblleft juggle\textquotedblright\ a hidden variable between
$\left\vert a\right\rangle $\ and $\left\vert b\right\rangle $, so that when
we inspect the hidden variable's history, both $\left\vert a\right\rangle
$\ and $\left\vert b\right\rangle $ are observed with high probability. \ The
difficulty is that this needs to work under \textit{any} indifferent
hidden-variable theory.

Next, Section \ref{SZK}\ combines the juggle subroutine with a technique of
Valiant and Vazirani \cite{vv}\ to prove that $\mathsf{SZK}\subseteq
\mathsf{DQP}$, from which it follows in particular that Graph Isomorphism\ and
Approximate Shortest Vector\ are in\ $\mathsf{DQP}$. \ Then Section
\ref{SEARCH}\ applies the juggle subroutine to search an $N$-item database in
$O\left(  N^{1/3}\right)  $\ queries, and also proves that this $N^{1/3}%
$\ bound is optimal. \ We conclude in Section \ref{OPEN}\ with some directions
for further research.

\section{The Computational Model\label{MODEL}}

We now explain our model of computation, building our way up to the complexity
class $\mathsf{DQP}$. \ Our starting point is the definition of
\textit{hidden-variable theory} that we gave in \cite{aaronson}. \ To recap
from that paper:\ for us a hidden-variable theory is simply a family of
functions $\left\{  S_{N}\right\}  _{N\in\left\{  1,2,\ldots\right\}  }$,
where each $S_{N}$\ maps an $N\times N$\ density matrix $\rho$\ and an
$N\times N$\ unitary matrix $U$\ onto an $N\times N$\ stochastic matrix
$S=S_{N}\left(  \rho,U\right)  $. \ In this paper, $\rho$\ will always be a
pure state of $l=\log_{2}N$ qubits. \ That is, $\rho=\left\vert \psi
\right\rangle \left\langle \psi\right\vert $\ where%
\[
\left\vert \psi\right\rangle =\sum_{x\in\left\{  0,1\right\}  ^{l}}\alpha
_{x}\left\vert x\right\rangle .
\]
What is essential is that $S$\ map the probability distribution induced by
measuring $\left\vert \psi\right\rangle $\ in the computational basis
$\left\{  \left\vert x\right\rangle \right\}  _{x\in\left\{  0,1\right\}
^{l}}$, onto the probability distribution induced by measuring $U\left\vert
\psi\right\rangle $\ in that same basis. \ More formally, let $\left(
M\right)  _{xy}$\ denote the entry in the $x^{th}$\ column and $y^{th}$\ row
of matrix $M$, and let%
\[
U\left\vert \psi\right\rangle =\sum_{x\in\left\{  0,1\right\}  ^{l}}\beta
_{x}\left\vert x\right\rangle .
\]
Then we require that for all $y\in\left\{  0,1\right\}  ^{l}$,%
\[
\sum_{x\in\left\{  0,1\right\}  ^{l}}\left(  S\right)  _{xy}\left\vert
\alpha_{x}\right\vert ^{2}=\left\vert \beta_{y}\right\vert ^{2}.
\]
It is clear that there are infinitely many theories satisfying the above
marginalization axiom; the simplest one is the \textit{product theory}
$\mathcal{PT}$, which sets $\left(  S\right)  _{xy}=\left\vert \beta
_{y}\right\vert ^{2}$\ for all $x,y$. \ To narrow down the choices, in
\cite{aaronson}\ we proposed seven additional axioms that we might want any
hidden-variable theory to satisfy. \ We then showed that, although not all of
the axioms can be satisfied simultaneously, two of the most important
ones---called indifference and robustness---\textit{can} be satisfied simultaneously.

Let us restate those two axioms in the present context.
\ \textit{Indifference} says that if $U$ is generalized block-diagonal (i.e. a
permutation of a block-diagonal matrix), then $S$ is also generalized
block-diagonal with the same block structure or some refinement thereof. \ So
in particular, if $\left\vert \psi\right\rangle $\ belongs to a tensor-product
Hilbert space $\mathcal{H}_{A}\otimes\mathcal{H}_{B}$, and if $U$ acts only on
$\mathcal{H}_{A}$\ (i.e. never maps a basis state $\left\vert x_{A}%
\right\rangle \otimes\left\vert x_{B}\right\rangle $\ to $\left\vert
y_{A}\right\rangle \otimes\left\vert y_{B}\right\rangle $\ where $x_{B}\neq
y_{B}$), then $S\left(  \left\vert \psi\right\rangle ,U\right)  $\ acts only
on $\mathcal{H}_{A}$\ as well. \ \textit{Robustness} says that $S$ is
insensitive to small perturbations of $\left\vert \psi\right\rangle $\ or $U$.
\ To make this intuition formal, we call a theory robust if for all $b>0$,
there exists $c>0$\ such that for all $l$, all pairs of states\ $\left\vert
\psi\right\rangle =\sum_{x\in\left\{  0,1\right\}  ^{l}}\alpha_{x}\left\vert
x\right\rangle $ and $\left\vert \widetilde{\psi}\right\rangle =\sum
_{x\in\left\{  0,1\right\}  ^{l}}\widetilde{\alpha}_{x}\left\vert
x\right\rangle $\ such that $\left\langle \psi|\widetilde{\psi}\right\rangle
\geq1-2^{-cl}$, and all $U$\ and $\widetilde{U}$ such that $\left\vert \left(
U\right)  _{xy}-\left(  \widetilde{U}\right)  _{xy}\right\vert \leq2^{-cl}%
$\ for all $x,y$, we have%
\[
\left\vert \left(  S\right)  _{xy}\left\vert \alpha_{x}\right\vert
^{2}-\left(  \widetilde{S}\right)  _{xy}\left\vert \widetilde{\alpha}%
_{x}\right\vert ^{2}\right\vert \leq2^{-bl}%
\]
for all $x,y$, where $S=S\left(  \left\vert \psi\right\rangle ,U\right)  $ and
$\widetilde{S}=S\left(  \left\vert \widetilde{\psi}\right\rangle
,\widetilde{U}\right)  $.

It is easy to show that the product theory $\mathcal{PT}$\ satisfies
robustness but not indifference. \ In \cite{aaronson},\ we analyzed three
other hidden-variable theories: the \textit{Dieks theory} $\mathcal{DT}$,
which satisfies indifference but not robustness; the \textit{flow theory}
$\mathcal{FT}$, which satisfies both indifference and robustness; and the
\textit{Schr\"{o}dinger theory} $\mathcal{ST}$, which satisfies indifference,
and which we conjecture satisfies robustness. \ The details of those theories
are mostly irrelevant for this paper. \ Indeed, our algorithms will work under
\textit{any} hidden-variable theory that satisfies the indifference axiom.
\ On the other hand, if we take into account that even in theory (let alone in
practice), a generic unitary cannot be represented exactly with a finite
universal gate set, only approximated arbitrarily well, then we also need the
robustness axiom. \ Thus, a key result from \cite{aaronson}\ that we rely on
is that there \textit{exists} a hidden-variable theory\ (namely $\mathcal{FT}%
$) satisfying both indifference and robustness.

Let a quantum computer have the initial state $\left\vert 0\right\rangle
^{\otimes l}$, and suppose we apply a sequence $\mathcal{U}=\left(
U_{1},\ldots,U_{T}\right)  $\ of unitary operations, each of which is
implemented by a polynomial-size quantum circuit. \ Then a \textit{history} of
a hidden variable through the computation\ is a sequence $H=\left(
v_{0},\ldots,v_{T}\right)  $ of basis states, where $v_{t}$\ is the variable's
value immediately after $U_{t}$\ is applied (thus $v_{0}=\left\vert
0\right\rangle ^{\otimes l}$). \ Given any hidden-variable theory
$\mathcal{T}$, we can obtain a probability distribution $\Omega\left(
\mathcal{U},\mathcal{T}\right)  $\ over histories by just applying
$\mathcal{T}$\ repeatedly, once for each $U_{t}$, to obtain the stochastic
matrices%
\[
S\left(  \left\vert 0\right\rangle ^{\otimes l},U_{1}\right)  ,~~S\left(
U_{1}\left\vert 0\right\rangle ^{\otimes l},U_{2}\right)  ,~~\ldots~~S\left(
U_{T-1}\cdots U_{1}\left\vert 0\right\rangle ^{\otimes l},U_{T}\right)  .
\]
Note that $\Omega\left(  \mathcal{U},\mathcal{T}\right)  $\ is a Markov
distribution; that is, each $v_{t}$\ is independent of the other $v_{i}$'s
conditioned on $v_{t-1}$\ and $v_{t+1}$. \ Admittedly, $\Omega\left(
\mathcal{U},\mathcal{T}\right)  $\ could depend on the precise way in which
the combined circuit $U_{T}\cdots U_{1}$\ is \textquotedblleft
sliced\textquotedblright\ into component circuits $U_{1},\ldots,U_{T}$. \ But
as we showed in \cite{aaronson}, such dependence on the granularity of
unitaries is unavoidable in any hidden-variable theory other than
$\mathcal{PT}$.

Given a hidden-variable theory $\mathcal{T}$, let $\mathcal{O}\left(
\mathcal{T}\right)  $\ be an oracle that takes as input a positive integer
$l$, and a sequence of quantum circuits $\mathcal{U}=\left(  U_{1}%
,\ldots,U_{T}\right)  $\ that act on $l$ qubits. \ Here each $U_{t}$\ is
specified by a sequence $\left(  g_{t,1},\ldots,g_{t,m\left(  t\right)
}\right)  $\ of gates chosen from some finite universal gate set $\mathcal{G}%
$. \ The oracle $\mathcal{O}\left(  \mathcal{T}\right)  $\ returns as output a
sample $\left(  v_{0},\ldots,v_{T}\right)  $\ from the history distribution
$\Omega\left(  \mathcal{U},\mathcal{T}\right)  $\ defined previously. \ Now
let $A$ be a deterministic classical Turing machine that is given oracle
access to $\mathcal{O}\left(  \mathcal{T}\right)  $. \ The machine $A$
receives an input $x$, makes a single oracle query to $\mathcal{O}\left(
\mathcal{T}\right)  $, then produces an output based on the response. \ We say
a set of strings $L$ is in $\mathsf{DQP}$\ if there exists an $A$ such that
for all sufficiently large $n$ and inputs $x\in\left\{  0,1\right\}  ^{n}$,
and all theories $\mathcal{T}$ satisfying the indifference and robustness
axioms, $A$ correctly decides whether $x\in L$\ with probability at least
$2/3$, in time polynomial in $n$.

Let us make some remarks about the above definition. \ There is no real
significance in our requirement that $A$ be deterministic and classical, and
that it be allowed only one query to $\mathcal{O}\left(  \mathcal{T}\right)
$. \ We made this choice only because it suffices for our upper bounds; it
might be interesting to consider the effects of other choices. \ However,
other aspects of the definition are not arbitrary. \ The order of quantifiers
matters; we want a single $A$ that works for \textit{any} hidden-variable
theory satisfying indifference and robustness. \ Also, we require $A$ to
succeed only for sufficiently large $n$ since by choosing a large enough
robustness parameter $c$, an adversary might easily make $A$ incorrect on a
finite number of instances.

\section{Basic Results\label{RESULTS}}

Having defined the complexity class $\mathsf{DQP}$, in this short section we
establish its most basic properties. \ First of all, it is immediate that
$\mathsf{BQP}\subseteq\mathsf{DQP}$; that is, sampling histories is at least
as powerful as standard quantum computation. \ For $v_{1}$, the first
hidden-variable value returned by $\mathcal{O}\left(  \mathcal{T}\right)  $,
can be seen as simply the result of applying a polynomial-size quantum circuit
$U_{1}$ to the initial state $\left\vert 0\right\rangle ^{\otimes l}$ and then
measuring in the standard basis.

A key further observation is the following.

\begin{proposition}
\label{anyuniv}Any universal gate set\ yields the same complexity class
$\mathsf{DQP}$. \ By universal, we mean here that any unitary matrix (real or
complex) can be approximated, without the need for ancilla qubits.
\end{proposition}

\begin{proof}
Let $\mathcal{G}$\ and $\widehat{\mathcal{G}}$\ be universal gate sets,
and\ let $U$ be a circuit made of $\operatorname*{poly}\left(  n\right)
$\ gates from $\mathcal{G}$. \ Then the Solovay-Kitaev Theorem
\cite{kitaev,nc}\ implies that we can approximate $U$\ to accuracy (say)
$2^{-nl}$\ by using $\operatorname*{poly}\left(  n,l\right)
=\operatorname*{poly}\left(  n\right)  $\ gates from $\widehat{\mathcal{G}}$,
which act on the same set of qubits as $U$ does. \ Furthermore, the
approximating circuit can be efficiently constructed. \ Now from the
definition of robustness, for all $\mathcal{T}$\ there exists a $c>0$\ such
that, if we approximate each $U_{t}\in\mathcal{U}$\ to accuracy $2^{-cl}$,
then the distribution\ over histories seen by $A$ is statistically
indistinguishable from what it would have been were the $U_{t}$'s represented
exactly. \ (This occurs when $b=3$\ for example.) \ Clearly $2^{-nl}\ll
2^{-cl}$\ for sufficiently large $n$.
\end{proof}

Unfortunately, the best upper bound on $\mathsf{DQP}$\ we have been able to
show is $\mathsf{DQP}\subseteq\mathsf{EXP}$; that is, any problem in
$\mathsf{DQP}$\ is solvable in deterministic exponential time. \ The proof is
trivial, but is the one place in the paper that\ relies on a specific
hidden-variable theory from \cite{aaronson}. \ Let $\mathcal{T}$\ be the flow
theory $\mathcal{FT}$, with the slight modification that we omit the step from
\cite{aaronson} of symmetrizing over all permutations of basis states. \ Then
by using the Ford-Fulkerson algorithm \cite{ff}, we can clearly construct the
requisite maximum flows in time polynomial in $2^{l}$\ (hence exponential in
$n$), and thereby calculate the probability of each possible history $\left(
v_{1},\ldots,v_{T}\right)  $\ to suitable precision. \ If we include the
symmetrization step, then we only know how to calculate these probabilities
in\ \textit{probabilistic} exponential time.

\section{The Juggle Subroutine\label{JUGGLE}}

This section presents a crucial subroutine that will be used in both
algorithms of this paper: the algorithm for simulating statistical zero
knowledge in Section \ref{SZK}, and the algorithm for search in $N^{1/3}%
$\ queries in Section \ref{SEARCH}. \ Given an $l$-qubit state $\left\vert
\psi\right\rangle =\left(  \left\vert a\right\rangle +\left\vert
b\right\rangle \right)  /\sqrt{2}$\ that is an equal superposition of two
unknown basis states, the goal of the juggle subroutine\ is to learn both $a$
and $b$. \ The name arises because our strategy will be to \textquotedblleft
juggle\textquotedblright\ a hidden variable, so that if it starts out at
$\left\vert a\right\rangle $\ then with non-negligible probability it
transitions to $\left\vert b\right\rangle $, and vice versa. \ Inspecting the
entire history of the hidden variable will then reveal both $a$ and $b$, as
desired. \ The difficulty is that we need a \textit{single} subroutine that
does this under \textit{all} hidden-variable theories satisfying the
indifference axiom---even theories that are designed specifically to thwart
such a subroutine. \ To meet this difficulty, we will apply a pair of
unitaries to $\left\vert \psi\right\rangle $ that force the hidden variable to
\textquotedblleft forget\textquotedblright\ whether it started at $\left\vert
a\right\rangle $\ or $\left\vert b\right\rangle $. \ We will then invert those
unitaries to return the state to $\left\vert \psi\right\rangle $, at which
point the hidden variable must be unequal to its initial value with
probability $1/2$.

We now give the subroutine. \ The first unitary, $U_{1}$, consists of Hadamard
gates on $l-1$\ qubits chosen uniformly at random, and the identity operation
on the remaining qubit, $i$. \ Next $U_{2}$ consists of a Hadamard gate on
qubit $i$. \ Finally $U_{3}$\ consists of Hadamard gates on all $l$ qubits.
\ Let $a=a_{1}\ldots a_{l}$\ and $b=b_{1}\ldots b_{l}$. \ Then since $a\neq
b$, we have $a_{i}\neq b_{i}$\ with probability at least $1/l$. \ Assuming
that occurs, the state%
\[
U_{1}\left\vert \psi\right\rangle =\frac{1}{2^{l/2}}\left(  \sum_{z\in\left\{
0,1\right\}  ^{l}~:~z_{i}=a_{i}}\left(  -1\right)  ^{a\cdot z-a_{i}z_{i}%
}\left\vert z\right\rangle +\sum_{z\in\left\{  0,1\right\}  ^{l}~:~z_{i}%
=b_{i}}\left(  -1\right)  ^{b\cdot z-b_{i}z_{i}}\left\vert z\right\rangle
\right)
\]
assigns nonzero amplitude to all $2^{l}$\ basis states. \ Then $U_{2}%
U_{1}\left\vert \psi\right\rangle $\ assigns nonzero amplitude to $2^{l-1}%
$\ basis states $\left\vert z\right\rangle $, namely those for which $a\cdot
z\equiv b\cdot z\left(  \operatorname{mod}2\right)  $. \ Finally $U_{3}%
U_{2}U_{1}\left\vert \psi\right\rangle =\left\vert \psi\right\rangle $.

Let $v_{t}$ be the value of the hidden variable after $U_{t}$\ is applied.
\ Then assuming $a_{i}\neq b_{i}$, we claim that $v_{3}$\ is independent of
$v_{0}$. \ So in particular, if $v_{0}=\left\vert a\right\rangle $\ then
$v_{3}=\left\vert b\right\rangle $\ with $1/2$ probability, and if
$v_{0}=\left\vert b\right\rangle $\ then $v_{3}=\left\vert a\right\rangle
$\ with $1/2$ probability. \ To see this, observe that when $U_{1}$\ is
applied, there is no interference between basis states $\left\vert
z\right\rangle $ such that $z_{i}=a_{i}$,\ and those such that $z_{i}=b_{i}$.
\ So by the indifference axiom, the probability mass at $\left\vert
a\right\rangle $\ must spread out evenly among all $2^{l-1}$\ basis states
that agree with $a$\ on the $i^{th}$\ bit, and similarly for the probability
mass at $\left\vert b\right\rangle $. \ Then after $U_{2}$\ is applied,
$v_{2}$\ can differ from $v_{1}$\ only on the $i^{th}$\ bit, again by the
indifference axiom. \ So each basis state\ of $U_{2}U_{1}\left\vert
\psi\right\rangle $\ must receive an equal contribution from probability mass
originating at $\left\vert a\right\rangle $, and probability mass originating
at $\left\vert b\right\rangle $. \ Therefore $v_{2}$\ is independent of
$v_{0}$, from which it follows that $v_{3}$\ is independent of $v_{0}$\ as well.

Unfortunately, the juggle subroutine only works with probability $1/\left(
2l\right)  $---for it requires that $a_{i}\neq b_{i}$,\ and even then,
inspecting the history $\left(  v_{0},v_{1},\ldots\right)  $\ only reveals
both $\left\vert a\right\rangle $\ and $\left\vert b\right\rangle $\ with
probability $1/2$. \ Furthermore, the definition of $\mathsf{DQP}$ does not
allow more than one call to the history oracle. \ However, all we need to do
is pack multiple subroutine calls into a single oracle call. \ That is, choose
$U_{4}$\ similarly to $U_{1}$\ (except with a different value of $i$), and set
$U_{5}=U_{2}$\ and $U_{6}=U_{3}$. \ Do the same with $U_{7}$, $U_{8}$, and
$U_{9}$, and so on. \ Since $U_{3},U_{6},U_{9},\ldots$\ all return the quantum
state to $\left\vert \psi\right\rangle $, the effect is that of multiple
independent juggle attempts. \ With $2l^{2}$\ attempts, we can make the
failure probability at most $\left(  1-1/\left(  2l\right)  \right)  ^{2l^{2}%
}<e^{-l}$.

As a final remark, it is easy to see that the juggle subroutine works equally
well with states of the form $\left\vert \psi\right\rangle =\left(  \left\vert
a\right\rangle -\left\vert b\right\rangle \right)  /\sqrt{2}$. \ This will
prove useful in Section \ref{SEARCH}.

\section{Simulating $\mathsf{SZK}$\label{SZK}}

Our goal is to show that $\mathsf{SZK}\subseteq\mathsf{DQP}$. \ Here
$\mathsf{SZK}$, or Statistical Zero Knowledge, was originally defined as the
class of all problems that possess a certain kind of \textquotedblleft
zero-knowledge proof protocol\textquotedblright---that is, a protocol between
an omniscient prover and a verifier, by which the verifier becomes convinced
of the answer to a problem, yet without learning anything else about the
problem. \ However, for our purposes this cryptographic definition of
$\mathsf{SZK}$\ is irrelevant. \ For Sahai and Vadhan \cite{sv}\ have given an
alternate and much simpler characterization: a problem is in $\mathsf{SZK}$ if
and only if it can be reduced to a problem called Statistical Difference,
which involves deciding whether two probability distributions are close or far.

More formally, let $P_{0}$\ and $P_{1}$\ be functions that map $n$-bit strings
to $n$-bit strings, and that are specified by classical polynomial-time
algorithms. \ Let $\Lambda_{0}$ and $\Lambda_{1}$\ be the probability
distributions over $P_{0}\left(  x\right)  $\ and $P_{1}\left(  x\right)
$\ respectively, if $x\in\left\{  0,1\right\}  ^{n}$ is chosen uniformly at
random. \ Then the problem is to decide whether $\left\Vert \Lambda
_{0}-\Lambda_{1}\right\Vert $\ is less than $1/3$\ or greater than $2/3$,
given that one of these is the case. \ Here%
\[
\left\Vert \Lambda_{0}-\Lambda_{1}\right\Vert =\frac{1}{2}\sum_{y\in\left\{
0,1\right\}  ^{n}}\left\vert \Pr_{x\in\left\{  0,1\right\}  ^{n}}\left[
P_{0}\left(  x\right)  =y\right]  -\Pr_{x\in\left\{  0,1\right\}  ^{n}}\left[
P_{1}\left(  x\right)  =y\right]  \right\vert
\]
is the variation distance between $\Lambda_{0}$ and $\Lambda_{1}$.

To illustrate, let us show that Graph Isomorphism is in $\mathsf{SZK}$.
\ Given two graphs $G_{0}$\ and $G_{1}$, take $\Lambda_{0}$\ to be the uniform
distribution over all permutations of $G_{0}$, and $\Lambda_{1}$\ to be
uniform over all permutations of $G_{1}$. \ This way, if $G_{0}$\ and $G_{1}%
$\ are isomorphic, then $\Lambda_{0}$ and $\Lambda_{1}$\ will be identical, so
$\left\Vert \Lambda_{0}-\Lambda_{1}\right\Vert =0$. \ On the other hand, if
$G_{0}$\ and $G_{1}$\ are non-isomorphic, then $\Lambda_{0}$ and $\Lambda_{1}%
$\ will be perfectly distinguishable, so $\left\Vert \Lambda_{0}-\Lambda
_{1}\right\Vert =1$. \ Since $\Lambda_{0}$\ and $\Lambda_{1}$ are clearly
samplable by polynomial-time algorithms, it follows that any instance of Graph
Isomorphism\ can be expressed as an instance of Statistical Difference. \ For
a proof that Approximate Shortest Vector is in $\mathsf{SZK}$, we refer the
reader to Aharonov and Ta-Shma \cite{at}.

Our proof will use the following \textquotedblleft amplification
lemma\textquotedblright\ from \cite{sv}:\footnote{Note that in this lemma, the
constants $1/3$\ and $2/3$ are not arbitrary; it is important for technical
reasons that $\left(  2/3\right)  ^{2}>1/3$.}

\begin{lemma}
[Sahai and Vadhan]\label{amp}Given efficiently-samplable distributions
$\Lambda_{0}$\ and $\Lambda_{1}$, we can construct new efficiently-samplable
distributions $\Lambda_{0}^{\prime}$\ and $\Lambda_{1}^{\prime}$, such that if
$\left\Vert \Lambda_{0}-\Lambda_{1}\right\Vert \leq1/3$ then $\left\Vert
\Lambda_{0}^{\prime}-\Lambda_{1}^{\prime}\right\Vert \leq2^{-n}$, while if
$\left\Vert \Lambda_{0}-\Lambda_{1}\right\Vert \geq2/3$\ then $\left\Vert
\Lambda_{0}^{\prime}-\Lambda_{1}^{\prime}\right\Vert \geq1-2^{-n}$.
\end{lemma}

In particular, Lemma \ref{amp} means we can assume without loss of generality
that either $\left\Vert \Lambda_{0}-\Lambda_{1}\right\Vert \leq2^{-n^{c}}$\ or
$\left\Vert \Lambda_{0}-\Lambda_{1}\right\Vert \geq1-2^{-n^{c}}$\ for some
constant $c>0$.

Having covered the necessary facts about $\mathsf{SZK}$, we can now proceed to
the main result.

\begin{theorem}
\label{szk}$\mathsf{SZK}\subseteq\mathsf{DQP}$.
\end{theorem}

\begin{proof}
We show how to solve Statistical Difference\ by using a history oracle. \ For
simplicity, we start with the special case where $P_{0}$\ and $P_{1}$\ are
both one-to-one functions. \ In this case, the circuit sequence $\mathcal{U}%
$\ given to the history oracle does the following: it first prepares the state%
\[
\frac{1}{2^{\left(  n+1\right)  /2}}\sum_{b\in\left\{  0,1\right\}
,x\in\left\{  0,1\right\}  ^{n}}\left\vert b\right\rangle \left\vert
x\right\rangle \left\vert P_{b}\left(  x\right)  \right\rangle .
\]
It then applies the juggle subroutine to the joint state of the $\left\vert
b\right\rangle $ and $\left\vert x\right\rangle $ registers, taking $l=n+1$.
\ Notice that by the indifference axiom, the hidden variable will never
transition from one value of $P_{b}\left(  x\right)  $\ to another---exactly
as if we had\ \textit{measured} the third register in the standard basis.
\ All that matters is the reduced state $\left\vert \psi\right\rangle $ of the
first two registers, which has the form $\left(  \left\vert 0\right\rangle
\left\vert x_{0}\right\rangle +\left\vert 1\right\rangle \left\vert
x_{1}\right\rangle \right)  /\sqrt{2}$\ for some $x_{0},x_{1}$\ if $\left\Vert
\Lambda_{0}-\Lambda_{1}\right\Vert =0$, and $\left\vert b\right\rangle
\left\vert x\right\rangle $\ for some $b,x$\ if $\left\Vert \Lambda
_{0}-\Lambda_{1}\right\Vert =1$. \ We have already seen that the juggle
subroutine can distinguish these two cases: when the hidden-variable history
is inspected, it will contain two values of the $\left\vert b\right\rangle $
register in the former case, and only one value in the latter case. \ Also,
clearly the case $\left\Vert \Lambda_{0}-\Lambda_{1}\right\Vert \leq2^{-n^{c}%
}$\ is statistically indistinguishable from $\left\Vert \Lambda_{0}%
-\Lambda_{1}\right\Vert =0$\ with respect to the subroutine, and likewise
$\left\Vert \Lambda_{0}-\Lambda_{1}\right\Vert \geq1-2^{-n^{c}}$\ is
indistinguishable from $\left\Vert \Lambda_{0}-\Lambda_{1}\right\Vert =1$.

We now consider the general case, where $P_{0}$\ and $P_{1}$\ need not be
one-to-one. \ Our strategy is to reduce to the one-to-one case, by using a
well-known hashing technique of Valiant and Vazirani \cite{vv}.\ \ Let
$\mathcal{D}_{n,k}$\ be the uniform distribution over all affine functions
mapping $\left\{  0,1\right\}  ^{n}$\ to $\left\{  0,1\right\}  ^{k}$, where
we identify those sets with the finite fields $\mathbb{F}_{2}^{n}$\ and
$\mathbb{F}_{2}^{k}$ respectively. \ What Valiant and Vazirani showed is that,
for all subsets $A\subseteq\left\{  0,1\right\}  ^{n}$\ such that $2^{k-2}%
\leq\left\vert A\right\vert \leq2^{k-1}$, and all $s\in\left\{  0,1\right\}
^{k}$,%
\[
\Pr_{h\in\mathcal{D}_{n,k}}\left[  \left\vert A\cap h^{-1}\left(  s\right)
\right\vert =1\right]  \geq\frac{1}{8}.
\]
As a corollary, the expectation over $h\in\mathcal{D}_{n,k}$\ of%
\[
\left\vert \left\{  s\in\left\{  0,1\right\}  ^{k}:\left\vert A\cap
h^{-1}\left(  s\right)  \right\vert =1\right\}  \right\vert
\]
is at least $2^{k}/8$. \ It follows that, if $x$\ is drawn uniformly at random
from $A$, then%
\[
\Pr_{h,x}\left[  \left\vert A\cap h^{-1}\left(  h\left(  x\right)  \right)
\right\vert =1\right]  \geq\frac{2^{k}/8}{\left\vert A\right\vert }\geq
\frac{1}{4}.
\]
This immediately suggests the following algorithm for the many-to-one case.
\ Draw $k$\ uniformly at random from $\left\{  2,\ldots,n+1\right\}  $; then
draw $h_{0},h_{1}\in\mathcal{D}_{n,k}$. \ Have $\mathcal{U}$\ prepare the
state%
\[
\frac{1}{2^{\left(  n+1\right)  /2}}\sum_{b\in\left\{  0,1\right\}
,x\in\left\{  0,1\right\}  ^{n}}\left\vert b\right\rangle \left\vert
x\right\rangle \left\vert P_{b}\left(  x\right)  \right\rangle \left\vert
h_{b}\left(  x\right)  \right\rangle ,
\]
and then apply the juggle subroutine to the joint state of the $\left\vert
b\right\rangle $\ and $\left\vert x\right\rangle $\ registers, ignoring the
$\left\vert P_{b}\left(  x\right)  \right\rangle $\ and $\left\vert
h_{b}\left(  x\right)  \right\rangle $ registers as before.

Suppose $\left\Vert \Lambda_{0}-\Lambda_{1}\right\Vert =0$. \ Also, given a
value $s=P_{b}\left(  x\right)  $, let $A_{0}=P_{0}^{-1}\left(  s\right)
$\ and $A_{1}=P_{1}^{-1}\left(  s\right)  $, and suppose $2^{k-2}%
\leq\left\vert A_{0}\right\vert =\left\vert A_{1}\right\vert \leq2^{k-1}$.
\ Then%
\[
\Pr_{s,h_{0},h_{1}}\left[  \left\vert A_{0}\cap h_{0}^{-1}\left(  s\right)
\right\vert =1\wedge\left\vert A_{1}\cap h_{1}^{-1}\left(  s\right)
\right\vert =1\right]  \geq\left(  \frac{1}{4}\right)  ^{2},
\]
since the events $\left\vert A_{0}\cap h_{0}^{-1}\left(  s\right)  \right\vert
=1$\ and $\left\vert A_{1}\cap h_{1}^{-1}\left(  s\right)  \right\vert
=1$\ are independent of each other conditioned on $s$. \ Assuming both events
occur, as before the juggle subroutine will reveal both $\left\vert
0\right\rangle \left\vert x_{0}\right\rangle $\ and $\left\vert 1\right\rangle
\left\vert x_{1}\right\rangle $\ with high probability, where $x_{0}$\ and
$x_{1}$\ are the unique elements of $A_{0}\cap h_{0}^{-1}\left(  s\right)
$\ and $A_{1}\cap h_{1}^{-1}\left(  s\right)  $\ respectively. \ By contrast,
if $\left\Vert \Lambda_{0}-\Lambda_{1}\right\Vert =1$\ then only one value of
the $\left\vert b\right\rangle $\ register will ever be observed. \ Again,
replacing $\left\Vert \Lambda_{0}-\Lambda_{1}\right\Vert =0$\ by $\left\Vert
\Lambda_{0}-\Lambda_{1}\right\Vert \leq2^{-n^{c}}$, and $\left\Vert
\Lambda_{0}-\Lambda_{1}\right\Vert =1$\ by $\left\Vert \Lambda_{0}-\Lambda
_{1}\right\Vert \geq1-2^{-n^{c}}$, can have only a negligible effect on the
history distribution.

Of course, the probability that the correct value of $k$ is chosen, and that
$A_{0}\cap h_{0}^{-1}\left(  s\right)  $\ and $A_{1}\cap h_{1}^{-1}\left(
s\right)  $\ both have a unique element, could be as low as $1/\left(
16n\right)  $. \ To deal with this, we simply increase the number of calls to
the juggle subroutine by an $O\left(  n\right)  $ factor, drawing new values
of $k,h_{0},h_{1}$ for each call. \ We pack multiple subroutine calls into a
single oracle call as described in Section \ref{JUGGLE}, except that now we
uncompute the entire state (returning it to $\left\vert 0\cdots0\right\rangle
$) and then recompute it between subroutine calls. \ A final remark: since the
algorithm that calls the history oracle is deterministic, we \textquotedblleft
draw\textquotedblright\ new values of $k,h_{0},h_{1}$ by having $\mathcal{U}%
$\ prepare a uniform superposition over all possible values. \ The
indifference axiom justifies this procedure, by guaranteeing that within
each\ call to the juggle subroutine, the hidden-variable values of $k$,
$h_{0}$, and $h_{1}$\ remain constant.
\end{proof}

Let us end this section with some brief remarks about the oracle result of
\cite{aarcol}. \ Given a function $g:\left\{  0,1\right\}  ^{n}\rightarrow
\left\{  0,1\right\}  ^{n}$, the \textit{collision problem} is to decide
whether $g$ is one-to-one or two-to-one, given that one of these is the case.
\ The question is, how many queries to $g$ are needed to solve this problem
(where a query just returns $g\left(  x\right)  $\ given $x$)? \ It is not
hard to see that $\Theta\left(  2^{n/2}\right)  $\ queries are necessary and
sufficient for classical randomized algorithms. \ What we showed in
\cite{aarcol}\ is that $\Omega\left(  2^{n/5}\right)  $\ queries are needed by
any \textit{quantum} algorithm as well. \ Subsequently Shi \cite{shi}\ managed
to improve the quantum lower bound to $\Omega\left(  2^{n/3}\right)  $
queries, thereby matching an upper bound of Brassard, H\o yer, and Tapp
\cite{bht}. \ On the other hand, the collision problem is easily reducible to
the Statistical Difference problem, and is therefore solvable in polynomial
time by sampling histories. \ This is the essence of the statement that
$\mathsf{BQP}\neq\mathsf{DQP}$\ relative to an oracle.

\section{Search in $N^{1/3}$ Queries\label{SEARCH}}

Given a Boolean function $f:\left\{  0,1\right\}  ^{n}\rightarrow\left\{
0,1\right\}  $, the database search problem is simply to find a string
$x$\ such that $f\left(  x\right)  =1$. \ We can assume without loss of
generality that this \textquotedblleft marked item\textquotedblright\ $x$ is
unique.\footnote{For if there are multiple marked items, then we can reduce to
the unique marked item case by using the Valiant-Vazirani hashing technique
described in Theorem \ref{szk}.} \ We want to find it using as few queries to
$f$ as possible, where a query returns $f\left(  y\right)  $\ given $y$.

Let $N=2^{n}$. \ Then classically, of course, $\Theta\left(  N\right)
$\ queries are necessary and sufficient. \ By querying $f$ in superposition,
Grover's algorithm \cite{grover}\ finds $x$ using $O\left(  N^{1/2}\right)
$\ queries, together with $\widetilde{O}\left(  N^{1/2}\right)  $\ auxiliary
computation steps (here the $\widetilde{O}$\ hides a factor of the form
$\left(  \log N\right)  ^{c}$). \ Bennett et al. \cite{bbbv}\ showed that any
quantum algorithm needs $\Omega\left(  N^{1/2}\right)  $\ queries.

In this section, we show how to find the marked item by sampling histories,
using only $O\left(  N^{1/3}\right)  $\ queries and $\widetilde{O}\left(
N^{1/3}\right)  $\ computation steps. \ Formally, the model is as
follows.\ \ Each of the quantum circuits $U_{1},\ldots,U_{T}$\ that algorithm
$A$ gives to the history oracle $\mathcal{O}\left(  \mathcal{T}\right)  $ is
now able to query $f$. \ Suppose $U_{t}$ makes $q_{t}$\ queries to $f$; then
the total number of queries made by $A$ is defined to be $Q=q_{1}+\cdots
+q_{T}$. \ The total number of \textit{computation} steps is at least the
number of steps required to write down $U_{1},\ldots,U_{T}$, but could be greater.

\begin{theorem}
\label{searchthm}In the $\mathsf{DQP}$ model, we can search a database of $N$
items for a unique marked item using $O\left(  N^{1/3}\right)  $ queries and
$\widetilde{O}\left(  N^{1/3}\right)  $\ computation steps.
\end{theorem}

\begin{proof}
Assume without loss of generality that $N=2^{n}$ with $n|3$, and that each
database item is labeled by an $n$-bit string. \ Let $x\in\left\{
0,1\right\}  ^{n}$ be the label of the unique marked item. \ Then the sequence
of quantum circuits $\mathcal{U}$ does the following: it first runs $O\left(
2^{n/3}\right)  $\ iterations of Grover's algorithm, in order to produce the
$n$-qubit\ state $\alpha\left\vert x\right\rangle +\beta\sum_{y\in\left\{
0,1\right\}  ^{n}}\left\vert y\right\rangle $, where%
\begin{align*}
\alpha &  =\sqrt{\frac{1}{2^{n/3}+2^{-n/3+1}+1}},\\
\beta &  =2^{-n/3}\alpha
\end{align*}
(one can check that this state is normalized). \ Next $\mathcal{U}$ applies
Hadamard gates to the first $n/3$\ qubits. \ This yields the state%
\[
2^{-n/6}\alpha\sum_{y\in\left\{  0,1\right\}  ^{n/3}}\left(  -1\right)
^{x_{A}\cdot y}\left\vert y\right\rangle \left\vert x_{B}\right\rangle
+2^{n/6}\beta\sum_{z\in\left\{  0,1\right\}  ^{2n/3}}\left\vert 0\right\rangle
^{\otimes n/3}\left\vert z\right\rangle ,
\]
where $x_{A}$\ consists of the first $n/3$ bits of $x$, and $x_{B}$\ consists
of the remaining $2n/3$ bits. \ Let $Y$\ be the set of $2^{n/3}$ basis states
of the form $\left\vert y\right\rangle \left\vert x_{B}\right\rangle $, and
$Z$\ be the set of $2^{2n/3}$\ basis states of the form $\left\vert
0\right\rangle ^{\otimes n/3}\left\vert z\right\rangle $.

Notice that $2^{-n/6}\alpha=2^{n/6}\beta$. \ So with the sole exception of
$\left\vert 0\right\rangle ^{\otimes n/3}\left\vert x_{B}\right\rangle $
(which belongs to both $Y$ and $Z$),\ the \textquotedblleft
marked\textquotedblright\ basis states in $Y$\ have the same amplitude as the
\textquotedblleft unmarked\textquotedblright\ basis states in $Z$. \ This is
what we wanted. \ Notice also that, if we manage to find any $\left\vert
y\right\rangle \left\vert x_{B}\right\rangle \in Y$, then we can find $x$
itself using $2^{n/3}$ further classical queries: simply test all\ possible
strings that end in $x_{B}$. \ Thus, the goal of our algorithm will be to
cause the hidden variable to visit an element of $Y$, so that inspecting the
variable's history reveals that element.

As in Theorem \ref{szk}, the tools that we need are the juggle subroutine, and
a way of reducing many basis states to two. \ Let $s$ be drawn uniformly at
random from $\left\{  0,1\right\}  ^{n/3}$. \ Then $\mathcal{U}$\ appends a
third register to $\left\vert \phi\right\rangle $, and sets it equal to
$\left\vert z\right\rangle $\ if the first two registers have the form
$\left\vert 0\right\rangle ^{\otimes n/3}\left\vert z\right\rangle $, or to
$\left\vert s,y\right\rangle $\ if they have the form $\left\vert
y\right\rangle \left\vert x_{B}\right\rangle $. \ Disregarding the basis state
$\left\vert 0\right\rangle ^{\otimes n/3}\left\vert x_{B}\right\rangle $ for
convenience, the result is%
\[
2^{-n/6}\alpha\left(  \sum_{y\in\left\{  0,1\right\}  ^{n/3}}\left(
-1\right)  ^{x_{A}\cdot y}\left\vert y\right\rangle \left\vert x_{B}%
\right\rangle \left\vert s,y\right\rangle +\sum_{z\in\left\{  0,1\right\}
^{2n/3}}\left\vert 0\right\rangle ^{\otimes n/3}\left\vert z\right\rangle
\left\vert z\right\rangle \right)  .
\]
Next $\mathcal{U}$\ applies the juggle subroutine to the joint state of the
first two registers. \ Suppose the hidden-variable value has the form
$\left\vert 0\right\rangle ^{\otimes n/3}\left\vert z\right\rangle \left\vert
z\right\rangle $ (that is, lies outside $Y$). \ Then with probability
$2^{-n/3}$\ over $s$, the first $n/3$\ bits of $z$ are equal to $s$. \ Suppose
this event occurs. \ Then conditioned on the third register being $\left\vert
z\right\rangle $, the reduced state of the first two registers is%
\[
\frac{\left(  -1\right)  ^{x_{A}\cdot z_{B}}\left\vert z_{B}\right\rangle
\left\vert x_{B}\right\rangle +\left\vert 0\right\rangle ^{\otimes
n/3}\left\vert z\right\rangle }{\sqrt{2}},
\]
where $z_{B}$\ consists of the last $n/3$\ bits of $z$. \ So it follows from
Section \ref{JUGGLE}\ that with\ probability $\Omega\left(  1/n\right)  $, the
juggle subroutine will cause the hidden variable to transition from
$\left\vert 0\right\rangle ^{\otimes n/3}\left\vert z\right\rangle $\ to
$\left\vert z_{B}\right\rangle \left\vert x_{B}\right\rangle $,\ and hence
from $Z$ to $Y$.

The algorithm calls the juggle subroutine $\Theta\left(  2^{n/3}n\right)
=\Theta\left(  N^{1/3}\log N\right)  $\ times, drawing a new value of $s$ and
recomputing the third register after each call. \ Each call moves the hidden
variable from $Z$ to $Y$ with independent\ probability $\Omega\left(
2^{-n/3}/n\right)  $; therefore with high probability \textit{some} call does
so. \ Note that this juggling phase does not involve any database queries.
\ Also, as in Theorem \ref{szk}, \textquotedblleft drawing\textquotedblright%
\ $s$ really means preparing a uniform superposition over all possible $s$.
\ Finally, the probability that the hidden variable ever visits the basis
state $\left\vert 0\right\rangle ^{\otimes n/3}\left\vert x_{B}\right\rangle
$\ is exponentially small (by the union bound), which justifies our having
disregarded it.
\end{proof}

A curious feature of Theorem \ref{searchthm}\ is the tradeoff between queries
and computation steps. \ Suppose we had run $Q$ iterations of Grover's
algorithm, or in other words made $Q$ queries to $f$. \ Then provided
$Q\leq\sqrt{N}$, the marked state $\left\vert x\right\rangle $\ would have
occurred with probability $\Omega\left(  Q^{2}/N\right)  $, meaning that
$\widetilde{O}\left(  N/Q^{2}\right)  $\ calls to the juggle subroutine would
have been sufficient to find $x$. \ Of course, the choice of $Q$ that
minimizes $\max\left\{  Q,N/Q^{2}\right\}  $ is $Q=N^{1/3}$. \ On the other
hand, had we been willing to spend $\widetilde{O}\left(  N\right)
$\ computation steps, we could have found $x$\ with only a \textit{single}
query!\footnote{One should not make too much of this fact; one way to
interpret it is simply that the \textquotedblleft number of
queries\textquotedblright\ should be redefined as $Q+T$\ rather than $Q$.}
\ Thus, one might wonder whether some other algorithm could push the number of
queries below $N^{1/3}$, without simultaneously increasing the number of
computation steps. \ The following theorem rules out that possibility.

\begin{theorem}
\label{lowerbound}In the $\mathsf{DQP}$ model,\ $\Omega\left(  N^{1/3}\right)
$\ computation steps are needed to search an $N$-item database for a unique
marked item. \ As a consequence, there exists an oracle relative to which
$\mathsf{NP}\not \subset \mathsf{DQP}$; that is, $\mathsf{NP}$-complete
problems are not efficiently solvable by sampling histories.
\end{theorem}

\begin{proof}
Let $N=2^{n}$ and $f:\left\{  0,1\right\}  ^{n}\rightarrow\left\{
0,1\right\}  $. \ Given a sequence of quantum circuits $\mathcal{U}=\left(
U_{1},\ldots,U_{T}\right)  $ that query $f$, and assuming that $x\in\left\{
0,1\right\}  ^{n}$\ is the unique string such that $f\left(  x\right)  =1$,
let $\left\vert \psi_{t}\left(  x\right)  \right\rangle $\ be the quantum
state after $U_{t}$\ is applied but before $U_{t+1}$\ is. \ Then the
\textquotedblleft hybrid argument\textquotedblright\ of Bennett et al.
\cite{bbbv}\ implies that, by simply changing the location of the marked item
from $x$ to $x^{\ast}$, we can ensure that%
\[
\left\Vert \left\vert \psi_{t}\left(  x\right)  \right\rangle -\left\vert
\psi_{t}\left(  x^{\ast}\right)  \right\rangle \right\Vert =O\left(
\frac{Q_{t}^{2}}{N}\right)
\]
where $\left\Vert ~~\right\Vert $\ represents trace distance, and\ $Q_{t}$ is
the total number of queries made to $f$ by $U_{1},\ldots,U_{t}$. \ Therefore
$O\left(  Q_{t}^{2}/N\right)  $\ provides an upper bound on the probability of
noticing the $x\rightarrow x^{\ast}$ change\ by monitoring $v_{t}$, the value
of the hidden variable after $U_{t}$\ is applied. \ So by the union bound, the
probability of noticing the change by monitoring the entire history $\left(
v_{1},\ldots,v_{T}\right)  $\ is at most of order%
\[
\sum_{t=1}^{T}\frac{Q_{t}^{2}}{N}\leq\frac{TQ_{T}^{2}}{N}.
\]
This cannot be $\Omega\left(  1\right)  $ unless $T=\Omega\left(
N^{1/3}\right)  $\ or $Q_{T}=\Omega\left(  N^{1/3}\right)  $, either of which
implies an $\Omega\left(  N^{1/3}\right)  $\ lower bound on the total number
of steps.

To obtain an oracle relative to which $\mathsf{NP}\not \subset \mathsf{DQP}$,
we can now use a standard and well-known \textquotedblleft diagonalization
method\textquotedblright\ due to Baker, Gill, and Solovay \cite{bgs} to
construct an infinite sequence of exponentially hard search problems, such
that any $\mathsf{DQP}$\ machine fails on at least one of the problems,
whereas there exists an $\mathsf{NP}$\ machine that succeeds on all of them.
\ We omit the details.
\end{proof}

\section{Discussion\label{OPEN}}

Perhaps the most interesting problem left open by this paper is the
computational complexity of simulating Bohmian mechanics. \ We strongly
conjecture that this problem, like the hidden-variable problems we have seen,
is strictly harder than simulating an ordinary quantum computer. \ The trouble
is that Bohmian mechanics does not quite fit in our framework: as discussed in
\cite{aaronson}, we cannot have deterministic hidden-variable trajectories for
discrete degrees of freedom such as qubits. \ Even worse, Bohmian mechanics
violates the continuous analogue of the indifference axiom. \ On the other
hand, this means that by trying to implement (say) the juggle subroutine with
Bohmian trajectories, one might learn not only about Bohmian mechanics and its
relation to quantum computation, but also about how essential the indifference
axiom really is for our implementation.

On the computer science side, a key open problem is to show better upper
bounds on $\mathsf{DQP}$. \ Recall that we were only able to show
$\mathsf{DQP}\subseteq\mathsf{EXP}$, by giving a classical exponential-time
algorithm to simulate the flow theory $\mathcal{FT}$. \ Can we improve this to
(say) $\mathsf{DQP}\subseteq\mathsf{PSPACE}$? \ Clearly it would suffice to
give a $\mathsf{PSPACE}$ algorithm that computes the transition probabilities
for some theory $\mathcal{T}$\ satisfying the indifference and robustness
axioms. \ On the other hand, this might not be \textit{necessary}---that is,
there might be an indirect simulation method that does not work by computing
(or even sampling from) the distribution over histories. \ It would also be
nice to pin down the complexities of simulating specific hidden-variable
theories, such as $\mathcal{FT}$\ and $\mathcal{ST}$.

\section{Acknowledgments}

I thank Umesh Vazirani, Ronald de Wolf, and an anonymous reviewer for comments
on an earlier version of this paper; Antony Valentini and Rob Spekkens for
helpful discussions; and Andris Ambainis for correcting an ambiguity in the
definition of $\mathsf{DQP}$. \ Supported by an NSF Graduate Fellowship and by
DARPA grant F30602-01-2-0524.

\end{document}